\documentclass[aps, prb, 10pt, twocolumn, superscriptaddress, longbibliography, nobalancelastpage]{revtex4-1}  

\usepackage{graphicx}
\usepackage{graphics}
\usepackage{amsmath}
\usepackage{amssymb}
\usepackage{amsfonts}
\usepackage{dsfont}
\usepackage{braket}
\usepackage{color}
\usepackage{braket,slashed}
\usepackage[mathscr]{euscript}
\usepackage[colorlinks=true,linkcolor=red,urlcolor=magenta,citecolor=blue]{hyperref}
\usepackage{subfigure}
\usepackage{xfrac}
\usepackage{bm}
\usepackage{kantlipsum}
\usepackage{enumitem}
\usepackage{tikz}
\usepackage{geometry}
\usepackage{framed}
\usepackage[english]{babel}

\allowdisplaybreaks[1]

\newcommand{\pepo}[0]{TNO}

\usepackage{amsmath}
\usepackage{amssymb}
\usepackage{graphicx}
\usepackage{dsfont}
\usepackage{braket}
\usepackage{bm}
\usepackage{wrapfig}
\usepackage{kantlipsum}
\usepackage{mathtools}
\usepackage[colorlinks=true,linkcolor=red,urlcolor=magenta,citecolor=blue]{hyperref}
\usepackage{tikz}
\usepackage[english]{babel}

\begin{document}

\title{Residual entropies for three-dimensional frustrated spin systems with tensor networks}

\author{Laurens Vanderstraeten}
\email{laurens.vanderstraeten@ugent.be}
\affiliation{Department of Physics and Astronomy, University of Ghent, Krijgslaan 281, 9000 Gent, Belgium}

\author{Bram Vanhecke}
\affiliation{Department of Physics and Astronomy, University of Ghent, Krijgslaan 281, 9000 Gent, Belgium}

\author{Frank Verstraete}
\affiliation{Department of Physics and Astronomy, University of Ghent, Krijgslaan 281, 9000 Gent, Belgium}
\affiliation{Vienna Center for Quantum Science and Technology, Faculty of Physics, University of Vienna, Boltzmanngasse 5, 1090 Vienna, Austria}

\date{\today}

\begin{abstract}
We develop a technique for calculating three-dimensional classical partition functions using projected entangled-pair states (PEPS). Our method is based on variational PEPS optimization algorithms for two-dimensional quantum spin systems, and allows us to compute free energies directly in the thermodynamic limit. The main focus of this work is classical frustration in three-dimensional many-body systems leading to an extensive ground-state degeneracy. We provide high-accuracy results for the residual entropy of the dimer model on the cubic lattice, water-ice $I_h$ and water-ice $I_c$. In addition, we show that these systems are in a Coulomb phase by computing the dipolar form of the correlation functions. As a further benchmark of our methods, we calculate the critical temperature and exponents of the Ising model on the cubic lattice.
\end{abstract}

\maketitle

\section{Introduction}

Many-body physics provides three distinct avenues for exhibiting interesting collective phenomena. First of all, classical phases transitions and critical behaviour originate from the battle between energy and entropy, where temperature tunes a system between different phases. Second, nontrivial phenomena can occur at zero temperature due to frustration, leading to nontrivial correlations and an extensive residual entropy in the system. Finally, quantum mechanics provides a very natural way of introducing frustration due to non-commuting operators and the associated zero-temperature quantum fluctuations.
\par In the language of tensor networks, it is obvious that these three many-body problems are intimately related. Indeed, the partition function of classical spin systems at finite temperature and the residual entropy of frustrated systems at zero-temperature can be readily formulated as a tensor network, whereas tensor-network states provide an excellent variational parametrization of ground states of quantum lattice systems. Yet, it is mainly this third avenue that has been pursued. The density matrix renormalization group (DMRG) has been used with great success for simulating quantum spin chains \cite{White1992, Schollwock2011}, whereas uniform matrix product states (MPS) \cite{Vidal2007, Haegeman2011} and projected entangled-pair states (PEPS) \cite{Verstraete2004, Jordan2008} algorithms have been constructed to tackle the ground-state problem for quantum spin systems in one and two spatial dimensions directly in the thermodynamic limit.
\par With respect to the first two classes of collective phenomena, all tensor-network efforts have been focused on simulating two-dimensional classical systems. The main goal of this paper is to demonstrate that these tensor-network techniques can be extended to the study of frustrated classical spin systems in three spatial dimensions. In particular, we translate recently developed PEPS algorithms for two-dimensional quantum lattice systems \cite{Vanderstraeten2016} to the three-dimensional classical setting. This variational tensor-network approach to classical spin systems was pioneered by T. Nishino and his collaborators \cite{Nishino2000, Nishino2001, Gendiar2002, Gendiar2003, Gendiar2005}, and here we follow up on this approach by invoking state-of-the-art tensor-network techniques. 
\par The structure of the paper is as follows. First we show that systems in two dimensions, such as the dimer covering problem and residual entropy problem of frustrated Ising models, can be readily calculated using standard MPS methods. Next, we show our results for the three-dimensional dimer problem and water ice. We then go on to demonstrate that our method also works well on more conventional problems by calculating the critical temperature of the three-dimensional Ising model. Finally we elaborate on how to construct the necessary tensor network, how its contraction is reduced to solving a variational PEPS optimization and and how we efficiently solve this optimization.

\section{Residual entropy calculation}

\par The third law of thermodynamics states that the entropy at zero temperature should always be zero. However, there are systems that violate this law by having a ground state that is, at least classically, extensively degenerate. An important example of such a system is regular water-ice, which was famously studied by L. Pauling \cite{Pauling1935} in 1935. He assumed that all configurations of water that satisfy the Bernal-Fowler ice rules\footnote{The Bernal-Fowler ice rules are such that each oxygen is covalently bonded to two hydrogens, each oxygen forms two hydrogen-bonds to two oxygens and there exists exactly one hydrogen between a pair of neigbouring oxygens.} were valid ground states. This enabled him to  explain the disagreement between the calorimetric and spectroscopic values of the absolute entropy of water. From what was essentially a mean-field approach, he could also estimate the entropy per site to be $S=k_b \ln(3/2)$, which has proven surprisingly accurate. Since then much progress has been made and the fields of frustrated systems, spin liquids and spin glasses have all developed from this early research. Yet, standard methods such as Monte-Carlo sampling and series expansions have great difficulty calculating residual entropies. Indeed, a Monte-Carlo simulation has no direct access to the free energy for a given value of temperature, and instead a full temperature range has to be simulated to obtain the value down to temperature zero.
\begin{table*}
	\begin{tabular}{ | l | p{3.5cm} | p{3.5cm} | p{3.5cm} |}
		\hline
		& AF-Ising on kagome & AF-Ising on triangular & Dimers on square\\ \hline
		TN & 0.5018331646 & 0.3230659407 & 0.2915608913 \\ \hline
		MPS bond dimension & 10 & 250 & 250\\ \hline
		Exact & 0.5018331646 & 0.3230659669 & $G/\pi\approx$ 0.2915609040\\ \hline
	\end{tabular}
	\caption{The tensor-network results were obtained by means of a vumps implementation \cite{Fishman2017}. The first two exact results where obtained by numerical integration of the relevant integrals, whereas in the last result we have used the numerical value of Catalan's constant $G$ \cite{catalan}.}
	\label{table1}
\end{table*}
\par In that respect, a tensor-network approach is a lot more natural, because the central object is the partition function itself and a direct variational optimization of the free energy is possible. In two dimensions there already exist powerful tools for performing this optimization \cite{Haegeman2017, Fishman2017}, where the central idea is to find an MPS approximation for the leading eigenvector of the `row-to-row' transfer matrix; the leading eigenvalue then determines the free energy. Here the bond dimension $D$ of the MPS is the control parameter, and by choosing $D$ large enough an arbitrary precision can be reached with moderate computational resources. We illustrate this by calculating the residual entropy of the antiferromagnetic Ising model on the triangular and kagome lattices, and the entropy of dimers on the square lattice. As all these models have exact solutions due to Kasteleyn's theorem \cite{Kasteleyn1963}, we can compare our results in Table \ref{table1}.

\par The accuracy of these results serves as a motivation for moving to three dimensions. In that case, the idea is to approximate the leading eigenvector of the `plane-to-plane' transfer matrix as a PEPS. Again, the bond dimension $D$ of the PEPS is the control parameter, and we expect that our results converge as $D$ is increased. The variational optimization of the eigenvalue of the transfer matrix then yields a value for the free energy, and other observables can be straightforwardly computed once a PEPS approximation has been found. In Secs.~\ref{sec:part} and \ref{sec:var} we will explain the tensor-network method in full detail, here we elaborate on the results.
\par We have studied two types of ground-state degeneracy, which differ by the rule that defines the set of ground states: the dimer-covering rule and the ice rule (on two different lattices). These two rules are special in that they may also be interpreted as imposing a global $U(1)$ symmetry. This special property entails that the models we study are expected to be in a Coulomb phase \cite{Henley2010}, which are described through a divergence-free field $\vec{B}$ exhibiting the asymptotic dipolar form 
\begin{equation} \label{eq:dipolar}
\braket{B_i(\vec{x}) B_j(0)} = \frac{1}{4\pi K} \frac{3x_ix_j -  |\vec{x}|^2\delta_{ij}}{ |\vec{x}|^5}, 
\end{equation}
where $K$ is the characteristic stiffness.
\par As a first benchmark we consider the cubic lattice dimer model. As was shown in Ref.~\onlinecite{Huse2003} this model is indeed in a critical Coulomb phase, characterized by algebraic dipolar forms for the dimer-dimer correlation functions. In particular, one can define a field variable on each link
\begin{equation}
B_j(\vec{x}) = (-1)^{|\vec{x}|} \left( n_j(\vec{x}) - 1/6 \right),
\end{equation}
where $\vec{x}$ denotes the lattice site and $n_j(\vec{x})=1$ if there is a dimer on that site in the direction $j$. The rule that there is one dimer on each lattice site implies that $B$ is the lattice version of a divergence-free field $\nabla\cdot B=0$. This vector field is now expected to exhibit the above dipolar form for its correlation functions.
\par
\begin{figure}
	\includegraphics[width=\columnwidth]{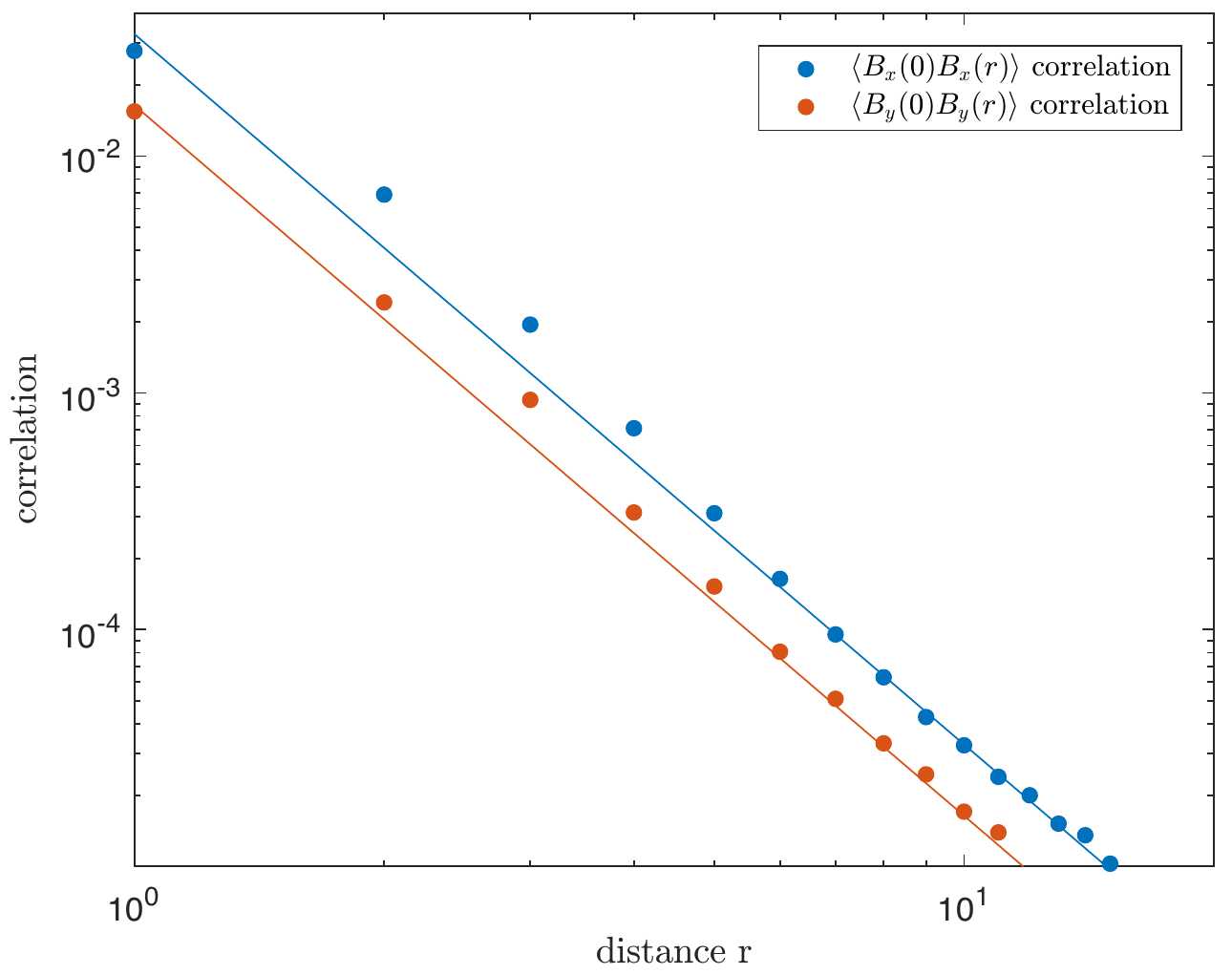}
	\caption{\label{fig:dimer}The correlation function $\braket{B_x(\vec{x}) B_x(0)}$ (blue) and $\braket{B_x(\vec{x}) B_y(0)}$ (red) along one of the horizontal axis, i.e. $\vec{x}=(r,0,0)$. The dots are the numerical results for an optimized $D=3$ PEPS, the lines are fits according to Eq.~\eqref{eq:dipolar} with the only free parameter $K=4.861$.}
\end{figure}

 Our PEPS simulation yields the following values for the dimer entropy: $0.4498238$ ($D=2$), $0.44988448$ ($D=3$) and $0.44988452$ ($D=4$). Correlation functions in this model are easily computed once an accurate PEPS approximation has been found. In Fig.~\ref{fig:dimer} we have plotted two correlation functions, showing a nice agreement with the above dipolar form. In particular, we find $K\approx4.9$, in agreement with the results reported in Ref.~\onlinecite{Huse2003}.

\par Secondly, we look at the residual entropy of ice in two different lattice structures. The first, ice $I_c$, has a lattice structure that is most easily understood as the familiar diamond lattice. The other is ice $I_h$, for which the crystal structure is less familiar (see Sec.~\ref{sec:part}). In both cases we have a crystal structure where each vertex has four edges and all edges form tetrahedral angles. The rules (ice rules) for allowed configurations of water-ice on such lattices are as follows: on each vertex resides an oxygen, the two hydrogens connected to this oxygen each lie on a different edge to that vertex and each edge has only one hydrogen on it. 

\par
\begin{figure}
	\includegraphics[width=\columnwidth]{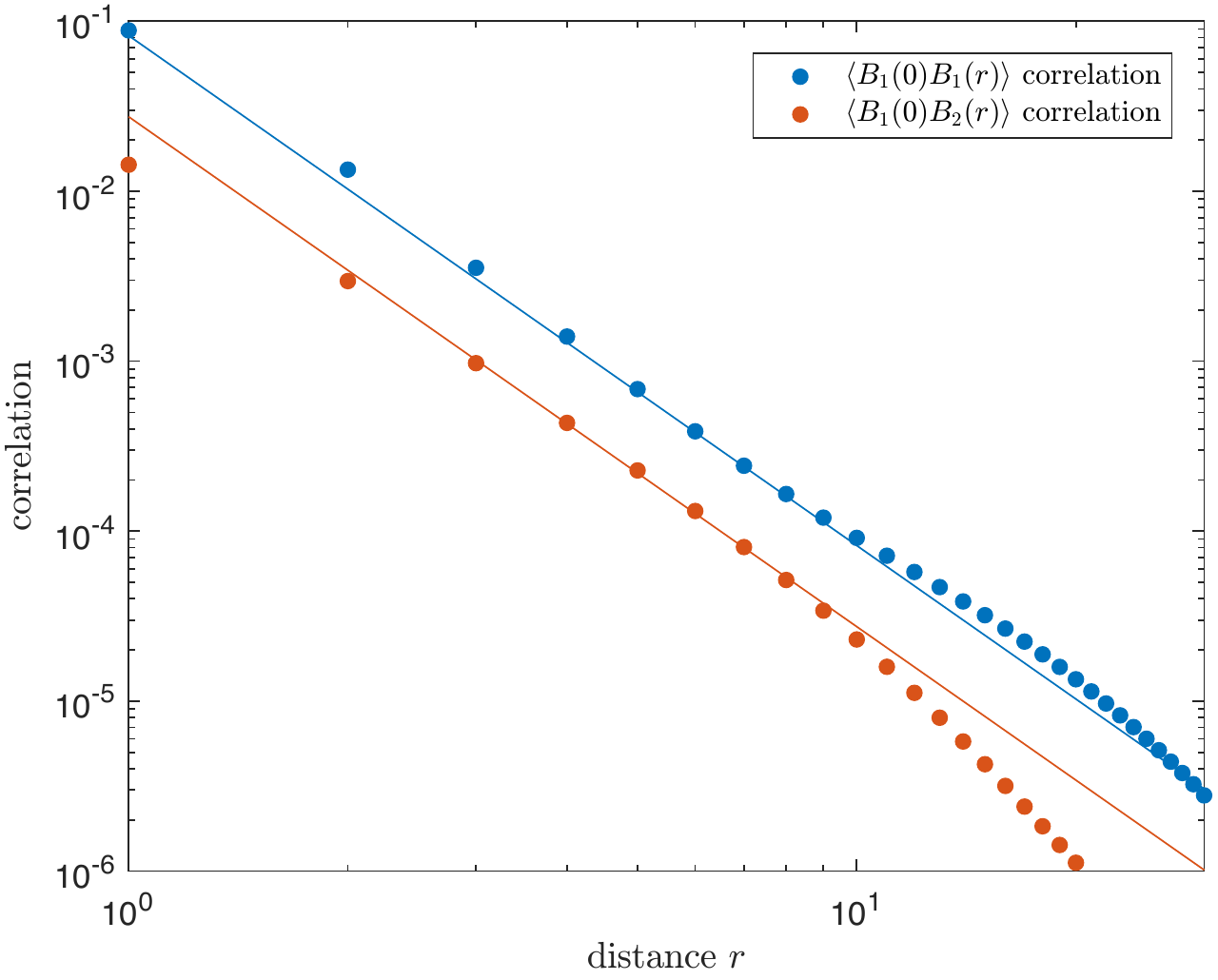}
	\caption{\label{fig:BfieldIce}The correlation function $\braket{B_1(\vec{x}) B_1(0)}$ (blue) and $\braket{B_1(\vec{x}) B_2(0)}$ (red) along one of the horizontal axis, i.e. $\vec{x}=(r,0,0)$. The dots are the numerical results for an optimized $D=3$ PEPS, the lines are fits according to Eq.~\eqref{eq:dipolar} with the only free parameter $K=0.967$. The directions $1$ and $2$ denote two directiones along 2 edges in the diamon lattice. Note that this correlation function should be identical for ice $I_h$ and ice $I_c$ since the direction $\vec{x}$ runs parallel to the \pepo{} (see Sec.~\ref{sec:part} for details).}
\end{figure}

 We ran our algorithm for PEPS of bond dimensions $D=2$, $D=3$ and $D=4$, and compare our results with the literature in Table \ref{table2}.
\begin{table}
	\begin{tabular}{ | l |l |}
		\hline
		Pauling\cite{Pauling1935} & 1.5 \\ \hline
		Nagle\cite{Nagle1966} & 1.506835(150) \\ \hline
		Berg et.al.\cite{Berg2007} & 1.50738(16) \\ \hline
		Kolafa \cite{Kolafa2014} & 1.5074660(36)\\ \hline
		D=2 & 1.50735\\ \hline
		D=3 & 1.507451 \\ \hline
		D=4 & 1.5074562 \\ \hline
	\end{tabular}
	\label{table2}
	\caption{Residual entropies from PEPS simulations compared to series expansion (Nagle) and Monte-Carlo simulations.}
\end{table}
We also calculated the equivalent correlation function of the dimer-dimer correlation function. Here the relevant field variable is 
\begin{equation}
B_j(\vec{x}) = 
\begin{cases}
+1,& \text{hydrogen on type-$A$ site}\\
-1, & \text{hydrogen on type-$B$ site}
\end{cases},
\end{equation}
where $\vec{x}$ denotes the lattice site and $j$ the direction; $A$ and $B$ denote the two sublattices (see Sec.~\ref{sec:part}). In Fig.~\ref{fig:BfieldIce} we show that we can nicely reproduce the dipolar form.
\par
\begin{figure}
	\includegraphics[width=\columnwidth]{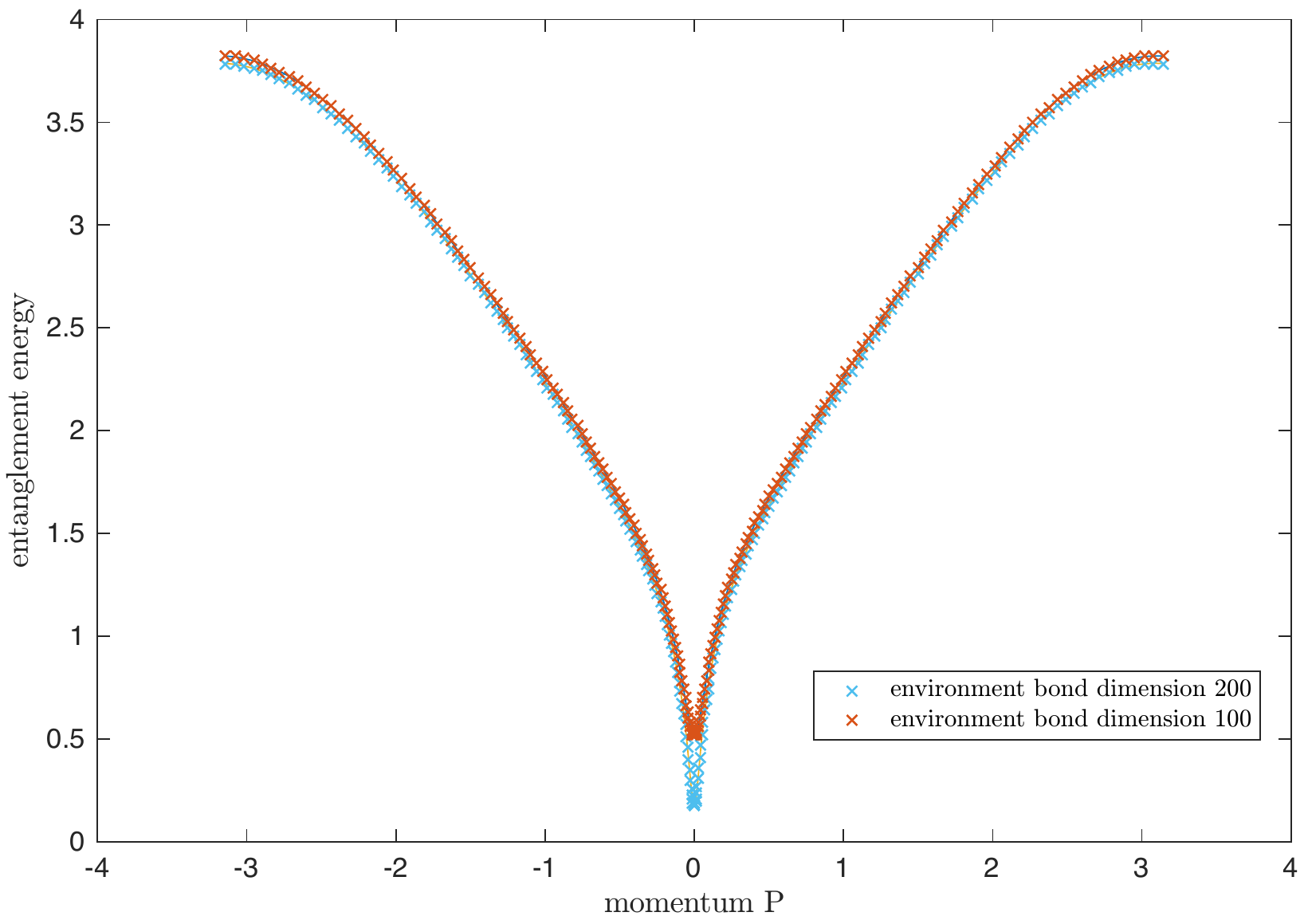}
	\caption{The entanglement spectrum of the $D=3$ PEPS for ice, which we computed using MPS quasiparticle ansatz as in Ref.~\onlinecite{Haegeman2016b}.}
	\label{fig:entanglement} 
\end{figure}

 As a matter of interest we also calculated the entanglement spectrum of the $D=3$ PEPS and plotted it in Fig.~\ref{fig:entanglement}. We find a symmetric (non-chiral) and gapless spectrum, and the dispersion seems to develop a cusp-like structure. This feature is reminiscent of dispersion relations of spin chains with power-law decaying interactions \cite{Vanderstraeten2018}, so we expect that the entanglement hamiltonian has similar properties. 

\section{Ising model}

\begin{figure*}
	\includegraphics[width=1.7\columnwidth]{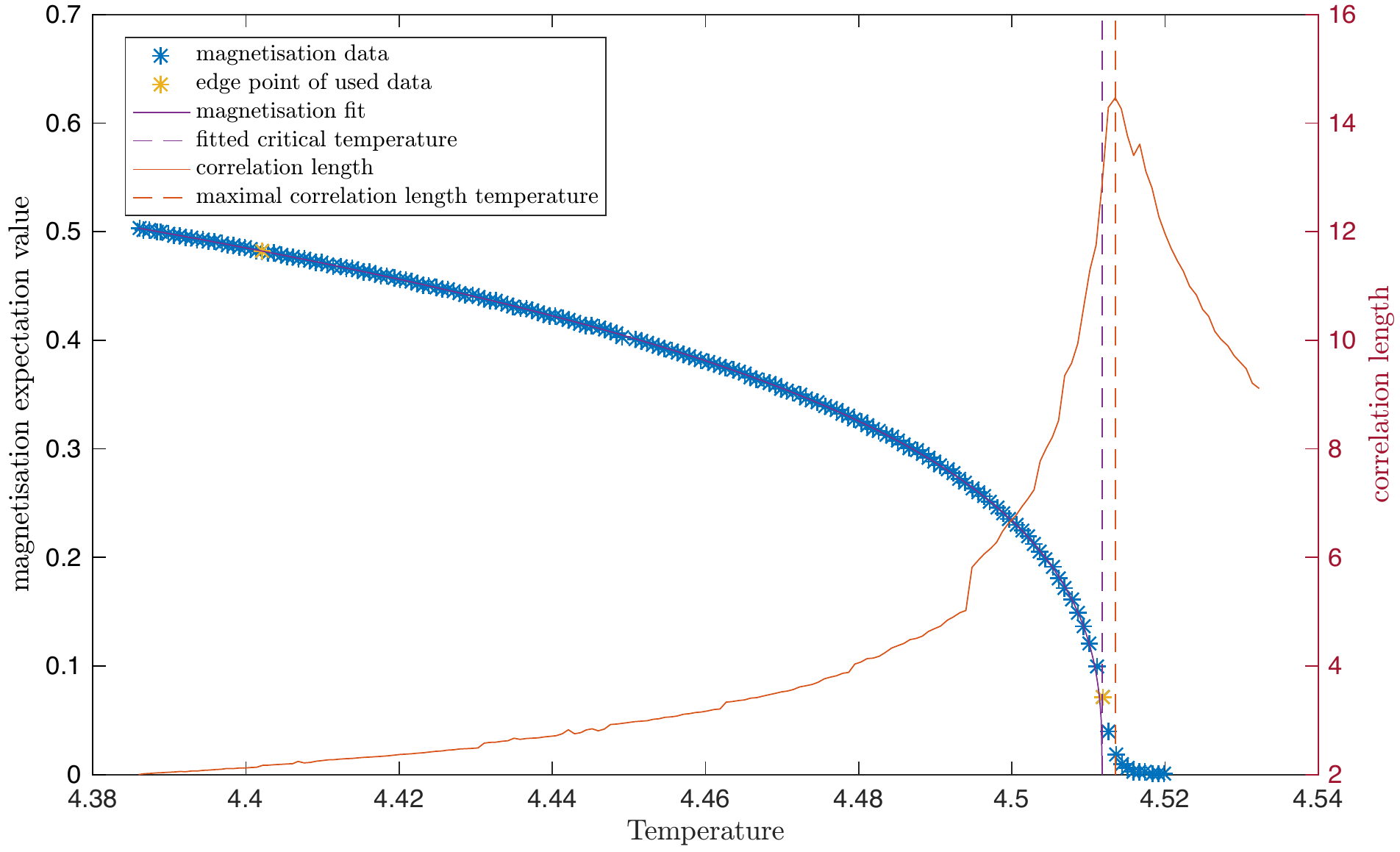}
	\caption{PEPS simulation of the classical Ising model with bond dimension $D=3$.}
	\label{fig:ising}
\end{figure*}

Finally we also test our method on the well-known Ising model. We have performed calculations for PEPS bond dimensions 2, 3 and 4. Because of the extreme sensitivity of the magnetization to small changes in the PEPS tensors, we had to converge our variational optimization to very high precision, i.e. the norm of the gradient was brought down to the order $5 \times 10^{-7}$. Optimizing a PEPS to this level of precision requires a lot of iterations in the numerical optimization, and appeared to be intractable for $D=4$ close to the critical point. For that reason, we obtain the best results for $D=3$. In Fig.~\ref{fig:ising} we have plotted our results for the Ising model. We nicely reproduce the sharp phase transition in the magnetization curve, but close to the critical point there is a rounding off due to the finite-$D$ approximation of the PEPS. In light of this finite-$D$ effect, we examine three methods for determining the critical temperature and compare them for the $D=3$ data.
\par First we use the correlation length for estimating the critical temperature. Although a finite-$D$ PEPS can exhibit critical correlations, it is expected that PEPS approximations for critical quantum Hamiltonians and transfer operators will always have a finite correlation length \cite{Rader2018, Corboz2018}. In order to obtain the correlation length for a given variational PEPS, we use an extrapolation technique proposed in Ref.~\onlinecite{Rams2018}. In Fig.~\ref{fig:ising} we have plotted the correlation length as a function of temperature, showing a strong maximum at a temperature slightly above the exact critical point. The maximum value of the correlation length we obtain is pretty high ($\xi_{\text{max}}\approx14$), and serves as a measure for the absolute error made around the critical point. Note that the correlation length is extremely sensitive to inaccuracies in the optimization -- even more so than the magnetization -- so it can fluctuate a bit despite our extreme precision.
\par Secondly, we try to fit the power-law behaviour to the magnetization curve, and extract an estimate for the critical temperature. We fit an algebraic function of the form $m(T)=T^\alpha(a_0+a_1T+a_2T^2)$ to the data. Note that we should not consider all the data points we calculated for the fit. At temperatures far from the critical temperature the fit, which is an expansion around the critical temperature, will start to fail. At temperatures close to critical our fit will also fail, this time due to the inherent limitations of our PEPS ansatz \cite{Rader2018, Corboz2018}. Therefore we look for an optimal window, not too close and not too far from the critical point to make the fit. We find this optimal window by optimizing over all possible windows with the confidence interval for the fitted critical temperature as a cost function.
\par As a PEPS typically favours breaking a symmetry (there is a bias towards lower-entanglement states), all these methods overestimate the critical temperature. The method that overestimates the least, and thus best approximates the true critical temperature, is the dynamical fitting technique as is illustrated in Fig.~\ref{fig:ising}. We find for the PEPS bond dimension 2, 3 and 4 the values $T_c=4.525222(33)$, $T_c=4.5118057(41)$ and $T_c=4.51195(65)$ respectively. As we noted above, the $D=4$ result lacks precision because we did not run the calculation all the way up to the critical temperature as the PEPS optimization became too costly. The critical exponent is also automatically extracted from this technique, we find $\beta=0.3688(6)$, $\beta=0.34620(77)$ and $\beta=0.327(11)$. These results are in good agreement with the high-accuracy results $T_c=4.511527$ \cite{Talapov1996} and $\beta=0.326419(3)$ \cite{Kos2016} found in the literature.

\section{Partition functions as tensor networks}
\label{sec:part}

We now explain in detail how the counting of dimer coverings and ice configurations, as well as classical partition functions, can be reformulated as the problem of contracting a tensor network. We note that both problems involve taking a sum of a number of terms that scales exponentially with the system size. The language of tensor networks now allows for an efficient representation of this exponentially large sum, where all information is encoded in a single tensor. This representation is extensive, in the sense that all observables scale naturally with system size and we can consider this network directly in the thermodynamic limit. As we will see in Sec.~\ref{sec:var}, all computations can be done in the thermodynamic limit as well, such that we never introduce finite-size effects.

\subsection{Dimer covering}

It is instructive to start with the simple example of the dimer-covering problem. We imagine placing dimers on a cubic lattice, where it is required that each site is part of exactly one dimer. The question we wish to answer is how many ways there are to perform such a covering. We can represent this number as the contraction of a cubic-lattice network of the same six-leg tensor, which is represented diagrammatically in Fig.~\ref{fig:cubic}. We find that the following six-leg tensor
\begin{equation}
    T^\text{dimer}_{i,j,k,l,m,n} =
    \begin{cases}
    1,& \text{one index has value 2}\\
    0,              & \text{otherwise}
    \end{cases}
\end{equation}
gives the right network. Indeed, this tensor can only contribute a factor one in each term of the whole sum if, and only if, exactly one of the bonds contains a dimer, otherwise it will contribute a factor zero and the entire term will drop away. Therefore, the full contraction of the network gives the number of possible dimer coverings, which scales exponentially with the number of tensors.
\par This tensor-network construction is completely generic; whenever one wishes to consider discrete degrees  on a certain lattice satisfying a local rule, a tensor network is easily constructed that counts the number of allowed configurations.

\subsection{Ice}

\par Extending the ideas used for constructing the tensor network for the dimer covering, we imagine making a network having the same lattice structure as either ice $I_h$ or ice $I_c$ and by analogy associate the index value 1 to `the hydrogen on this link is bound to the oxygen on this vertex' and value 2 to `the hydrogen is bound to the oxygen at this vertex'. This gives us following tensor:
\begin{equation}
T^\text{ice}_{i,j,k,l} =
\begin{cases}
1,& \text{two indices have value 2}\\
0,              & \text{otherwise}
\end{cases}.
\end{equation}
In addition we place a Pauli $X$ matrix on each edge, such that the hydrogen on each link is bound to one of the two oxygens. The resulting network gives you exactly the number of configurations satisfying the ice rules on that particular lattice. We can further simplify things by noticing that the lattice of ice $I_h$ is bipartite, and label all vertices by either A or B such that no A is ever connected to another A and neither are the B's. Next we contract all $X$-matrices with their adjacent A-labeled tensors. The tensors $T$ happen to have a $Z_2$ symmetry such that this operation leaves them unchanged, and so we end up with a tensor network that is made up of only tensors $T^\text{ice}$.

\par
\begin{figure}
	\includegraphics[width=0.8\columnwidth]{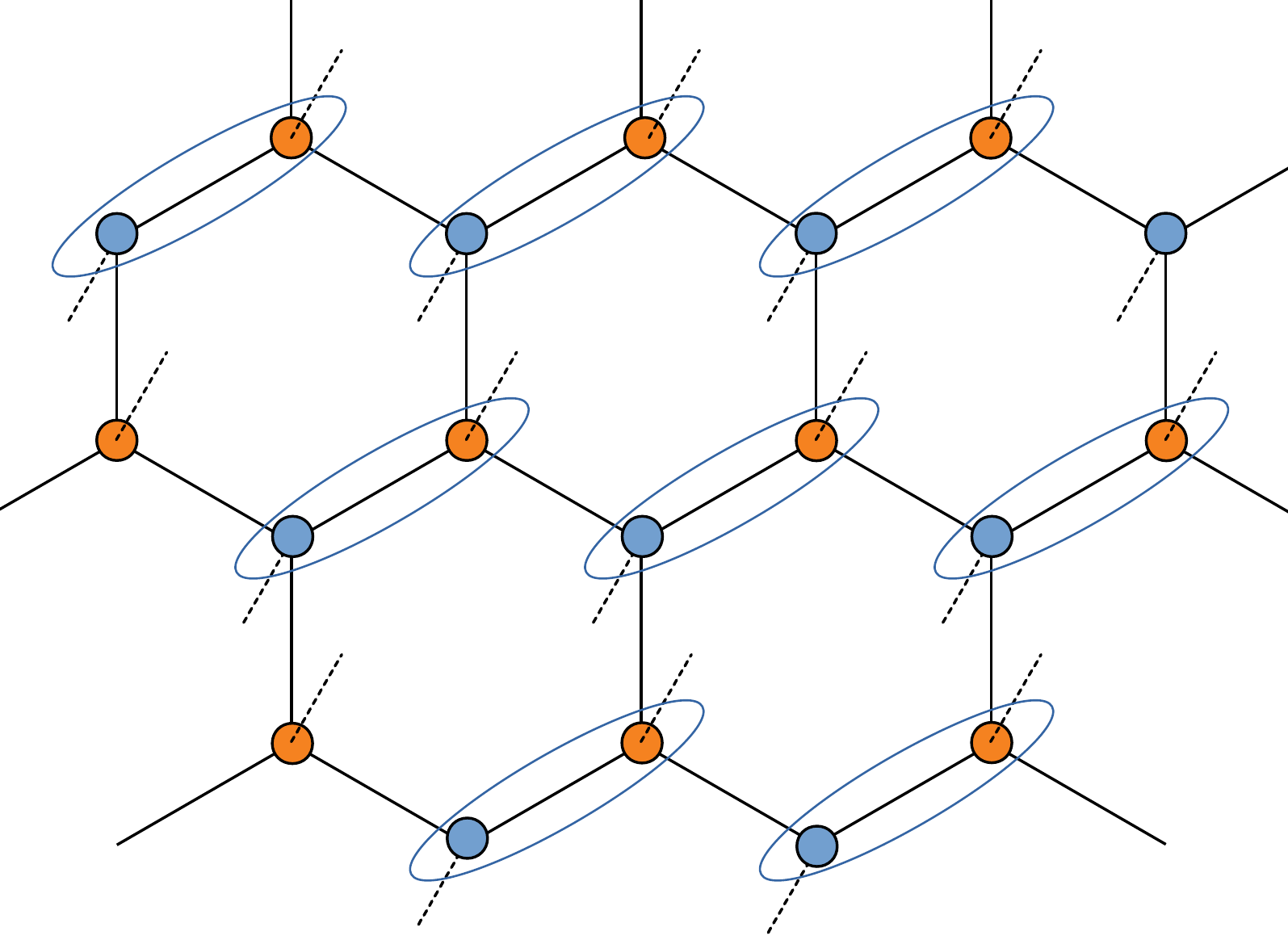}
	\caption{The diagrammatic representation of $A_{\text{hex}}$. The hexagonal lattice can be mapped to a square lattice by grouping pairs of tensors into a single tensor. The orange tensors represent the A-type tensors, the blue ones represent B-type. The square-lattice TNO for $I_c$ is obtained by stacking two of these, whereas $I_h$ is obtained by stacking $A_\text{hex}$ with its transpose. }
	\label{fig:hex}
\end{figure}

 For ice $I_c$ these tensors are arranged in a diamond-shaped network, but we can rewrite it as a cubic lattice problem as follows. The two-dimensional \pepo{} that generates the diamond lattice has the shape of a hexagonal lattice. Since this lattice is also bipartite, we can label each vertex by A or B such that no two A's or two B's are connected by an edge. On each A-type vertex we now imagine placing an index pointing up (the bra direction) and similarly we place an index pointing down (the ket direction) on all B-type vertices. Each vertex now has four edges connected to it and, in the lattices we will consider, these legs are organized in tetrahedral angles. We will refer to the resultant hexagonal \pepo{} as $A_{\text{hex}}$. The hexagonal lattice can be mapped to a square lattice by grouping pairs of tensors into a single tensor. For the diagrammatic representation of $A_{\text{hex}}$, see Fig.~\ref{fig:hex}.
\par In order to obtain the diamond lattice, one places two $A_{\text{hex}}$'s exactly on top of each other and then move the top $A_{\text{hex}}$ along one of the edges of the bottom $A_{\text{hex}}$ such that the A-type vertices of the bottom $A_{\text{hex}}$ sit right below the B-labeled vertices of the top $A_{\text{hex}}$. Then we contract the bottom-bra with the top-ket indices.

\par To construct the lattice of ice $I_h$ we can reuse $A_{\text{hex}}$, but this time we alternate $A_{\text{hex}}$ and $A_{\text{hex}}^T$ (the transpose of $A_{\text{hex}}$). Since the bra and ket indices are already aligned due to the transpose operation there is no need to move any of the TNOs relative to each other to connect them.

\subsection{Ising model}

\par
\begin{figure}
\includegraphics[width=0.6\columnwidth]{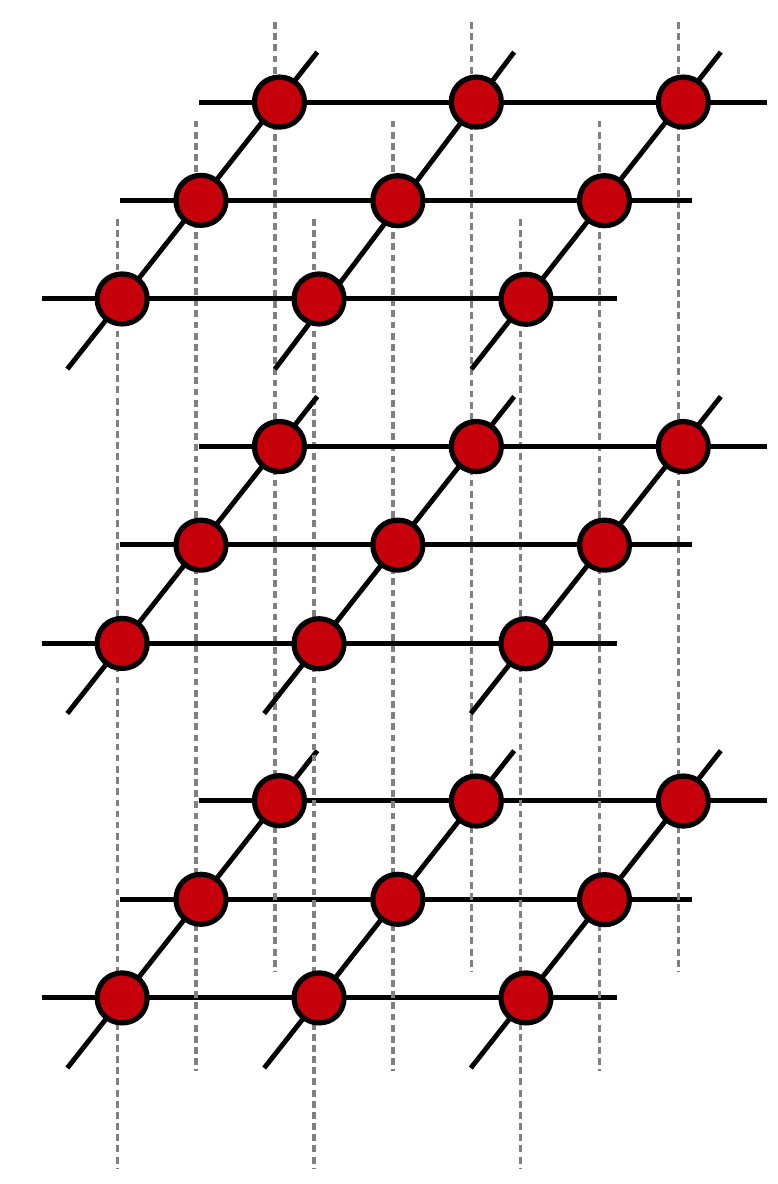}
\caption{\label{fig:cubic}Tensor network representation of a three-dimensional partition function. In this representation, the transfer matrix $T$ is clearly identified as a two-dimensional tensor-network operator.}
\end{figure}

 The partition function for a spin system with local interactions is similarly represented as a tensor network. For a nearest-neighbour interaction $H(s_1,s_2)$, the partition function is given by
\begin{equation}
\mathcal{Z}=\sum_{\{s\}}\prod_{\braket{ij}}e^{-\beta H(s_i,s_j)} .
\end{equation}
To construct the associated tensor network we can imagine first making a different one, the network that counts the number of allowed spin configurations. This can be achieved by placing following tensors on each vertex
\begin{equation}
T_{i,j,k,l,m,n} = 
\begin{cases}
1,& \text{if all idices are the same}\\
0,              & \text{otherwise}
\end{cases}.
\end{equation}
The index value 1 here means that the spin on this vertex is down, whereas the index value 2 corresponds to an up spin. As it stands, the network built from this $T$ would contract to 2, representing all spins up and all spins down. However, all spin configurations are in fact allowed, so all index values should be connected with each other, i.e. a 2x2 matrix of ones should be placed on all edges. This network would sum over all possible spin configurations, giving them all weight one. We are now left to add a Boltzmann weight to all nearest neighbours, weighing each configuration in the appropriate way. We can achieve this by replacing the matrix of ones on all edges by the following matrix:
\begin{equation}
t=
\begin{pmatrix}
e^{-\beta H(1,1)} & e^{-\beta H(1,2)}\\
e^{-\beta H(2,1)} & e^{-\beta H(2,2)}\\
\end{pmatrix}
=\begin{pmatrix}
e^{\beta} & e^{-\beta}\\
e^{-\beta} & e^{\beta}\\
\end{pmatrix}.
\end{equation}
In order to reduce this tensor network to the form of Fig.~\ref{fig:cubic}, we contract the square root of this matrix $t$ with each of the legs of tensor $T$ resulting in the final tensor $T^\text{Ising}$.

\begin{figure}
\includegraphics[width=0.8\columnwidth]{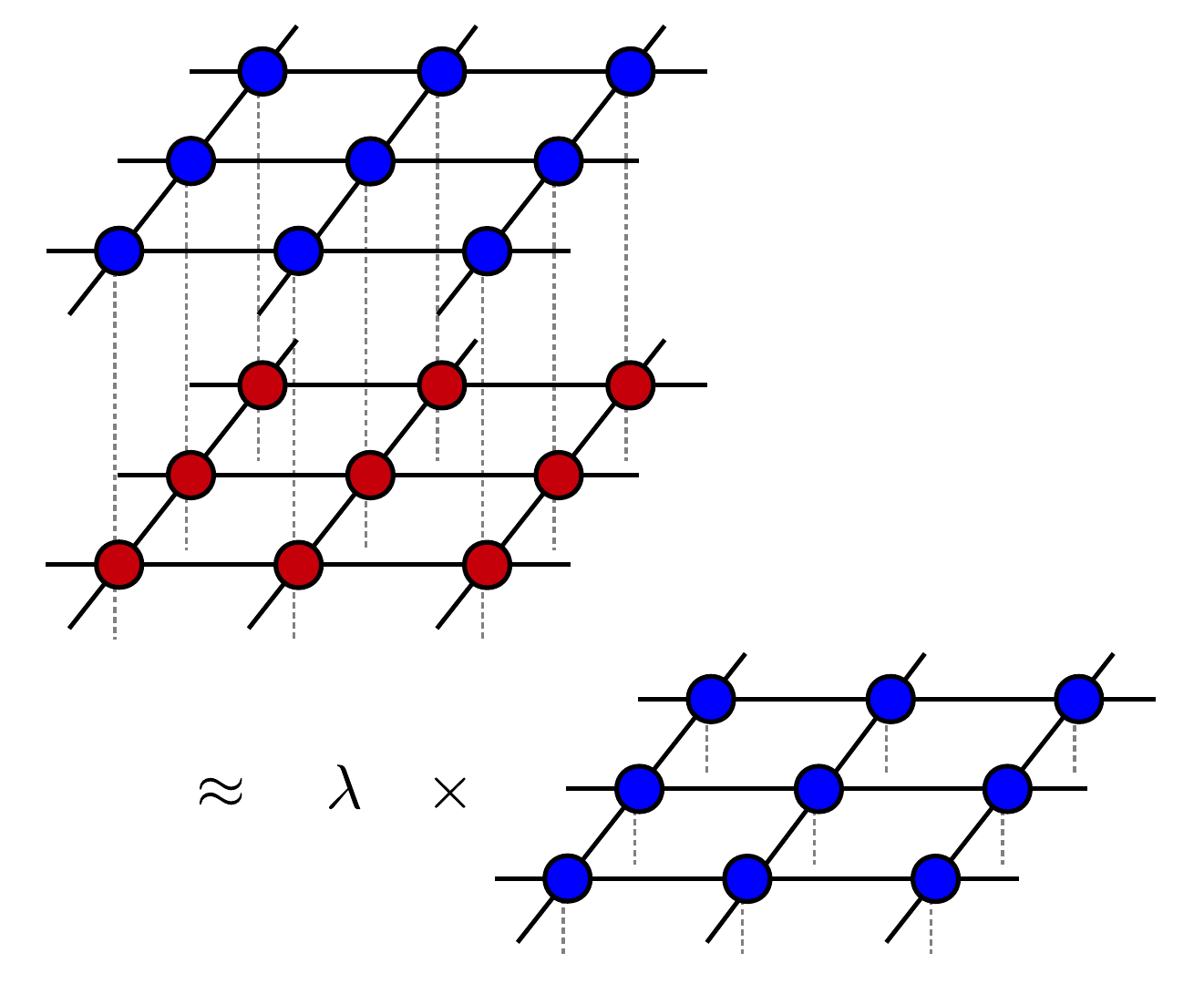}
\caption{\label{fig:peps_eig}Tensor network representation of the eigenvalue equation for the transfer matrix. The blue tensors make up a PEPS.}
\end{figure}

\section{Variational method}
\label{sec:var}

After having translated our problem to a tensor network, be it a classical partition function or a counting problem, we are now faced with the problem of contracting the network. From Fig.~\ref{fig:cubic} it is clear that, in the case of a cubic lattice, the transfer operator that generates the network, takes the form of a two-dimensional tensor network operator (TNO); it resembles a PEPS, but with physical degrees of freedom on both the top and bottom. The challenge of contracting the cubic network is thus to find the leading eigenvector of this transfer operator, i.e. solve the eigenvalue equation
\begin{equation}
T \ket{\Psi} = \lambda \ket{\Psi}.
\end{equation}
Here the eigenvalue $\lambda$ scales exponentially with the number of sites in the plane, such that the `free energy' $f=-\log\lambda$ is extensive in the system size. The basic idea of our algorithm is to approximate this leading eigenvector as a PEPS (see Fig.~\ref{fig:peps_eig}). If the transfer operator is hermitian we can find an optimal PEPS approximation variationally by optimizing the expectation value of the TNO with respect to the PEPS.
\par The basic tools for performing this variational optimization have already been developed in Ref.~\onlinecite{Vanderstraeten2016} (see also Ref.~\onlinecite{Corboz2016}). The situation differs in the fact that the operator whose largest eigenvalue we seek is not a Hamiltonian but rather a \pepo{}, which greatly simplifies the complexity of the algorithm as there is no need for channel environments \cite{Vanderstraeten2015}. The expression we wish to maximize is the following:
\begin{equation} \label{eq:costfunction}
f(A,\bar{A}) = - \log\left( \frac{\bra{\Psi(\bar{A})} T \ket{\Psi(A)}} {\braket{\Psi(\bar{A})|\Psi(A)}}\right),
\end{equation}
where $\ket{\Psi(A)}$ is a PEPS parametrized by a single tensor $A$ (see Fig.~\ref{fig:peps_eig}). The denominator of this expression is the familiar overlap of a PEPS with itself, whereas the numerator is the same but with the \pepo{} sandwiched between the bra- and ket-layer PEPS. In this work, we have applied the vumps algorithm \cite{Fishman2017} to compute both quantities. Following Ref.~\onlinecite{Vanderstraeten2016}, the derivative w.r.t. the PEPS degrees of freedom of $\lambda$ is given by
\begin{multline}
g = - 2 \times \left( \frac{\bra{\partial_{\bar{A}} \Psi(\bar{A})} T \ket{\Psi(A)}} {\bra{\Psi(\bar{A})} T \ket{\Psi(A)}} \right. \\ - \left. \frac{\braket{\partial_{\bar{A}} \Psi(\bar{A}) | \Psi(A)}} {\braket{\Psi(\bar{A})|\Psi(A)}} \right) .
\end{multline}
The second term is the one also encountered in Hamiltonian PEPS calculations\cite{Vanderstraeten2016}: it is a PEPS double-layer with one tensor in the bra missing. Similarly the first term is a triple-layer built from a sandwich of the \pepo{} between the PEPS and the conjugate PEPS, where again a PEPS tensor in the bra is missing. Both can be calculated by computing the effective one-site environment, again  using the vumps algorithm \cite{Fishman2017}.
\par These two quantities (objective function $f$ and gradient $g$) are then used to run a quasi-Newton optimization scheme. Though this algorithm is conceptually and implementationally very easy, the computational cost of running it can be high. If the network being contracted carries polynomial correlations, the double- and triple-layer will do so as well. This can cause the required MPS bond dimensions of the environments to quickly grow intractable. However, this only develops into a problem close to convergence, as the PEPS will only be truly critical if it is somehow the exact eigenvector of the critical TNO. Furthermore it is beneficial to adjust the MPS bond dimensions such that the precision with which one calculates the one-site environment is proportional to the norm of the gradient at that particular stage of the optimization. These factors combine to the general observation that low precision critical or non-critical calculations can be done quickly and reliably, but high precision calculations in critical models are hard.
\par Another subtlety of this technique is that it makes use of the variational characterization of the ground state of a Hermitian operator, and for non-Hermitian transfer TNOs this method is not guaranteed to work. Whereas the dimer-covering and Ising-model TNOs were real-symmetric, the square-lattice TNO that we have constructed for ice, is not. We can still run our algorithm but we must limit our search to the set of PEPS for which the \pepo{} is Hermitian. We do this by first noting that $A_{\text{hex}}$ may be tranformed into $A_{\text{hex}}^T$ by rotating $A_{\text{hex}}$ over 180 degrees around the center of one of its edges, as can be judged from Fig.~\ref{fig:hex} where interchanging the A- and B-labels is equivalent to a 180 degree rotation. Therefore, if we restrict to PEPS that are also invariant under this rotation, the variational optimization is indeed over a Hermition operator. Finite-size diagonalization of the \pepo{} with periodic boundary conditions confirms this assumption for the fixed point, and, due to the critical features of this model, we expect that rotational invariance is obeyed. As a consequence, our calculation for ice $I_h$ and ice $I_c$ will yield identical values for the residual entropy, a fact that is also observed to very high precision in Monte-Carlo simulations \cite{Kolafa2014}.

\section{Outlook}

In this paper we have described a generic method for contracting three-dimensional tensor networks, and applied this to study classical spin models and residual entropies in three dimensions. In particular, we have simulated the phase transition of the cubic-lattice Ising model, as well as the dimer-covering and water-ice problems. For the latter, we obtain excellent variational values for the residual entropy and we reproduce the critical Coulomb-phase correlations in these systems.
\par The results in this paper show that tensor-network methods can be applied to the simulation of classical statistical mechanics and frustration in three dimensions. The next step forward is taken by adding quantum mechanics to the picture, and investigating the interplay between these three origins of collective behaviour in many-body systems. In particular, the tensor networks that we have introduced to describe the extensive ground-state degeneracy on the classical level, can be straightforwardly extended to capture quantum states living in this exponentially large subspace -- we only need to add physical indices. Moreover, our methods allow us to compute expectation values for these three-dimensional quantum states and, a fortiori, make it possible to perform a variational optimization for the quantum-mechanical energy.
\par More generally, we expect that this paper provides the crucial stepping stone to lift the tensor-network simulation of quantum lattice systems to the three-dimensional world.

\section{Acknowledgements}

This work was supported by the Flemish Research Foundation, the Austrian Science Fund (ViCoM, FoQuS), the European Commission (QUTE 647905, ERQUAF 715861).

\bibliography{./bibliography}

\end{document}